\documentstyle[12pt,amssymb]{article}
\begin{document}
\input psfig
\def\today{December 31, 1998}
\def\gW{$\Omega_m$}
\def\gs{$\sigma_{8}$}
\def\gm{$\pm$}
\def\gi{$\sim$}
\def\gie{$\simeq$}
\def\gE{{\em et al.}}
\def\mlb{$\rm M/L_B$}
\def\ls{$\lesssim$}
\def\ss{$\gtrsim$}
\title{\bf Cosmology with Clusters of Galaxies}
\author{Neta A. Bahcall\footnote{\it neta@astro.princeton.edu}\\
Princeton University Observatory\\
Princeton, NJ 08544}
\maketitle
\begin{abstract}
Rich clusters of galaxies, the largest virialized systems known, place some
of the most powerful constraints on cosmology.  I discuss below the use of
clusters of galaxies in addressing two fundamental questions: What is the 
mass-density of the universe? and how is the mass distributed?  I show that 
several independent methods utilizing clusters of galaxies---cluster dynamics
and mass-to-light ratio, baryon fractions in clusters, and cluster evolution---
all indicate the same robust result:  the mass-density of the universe is low, 
\gW\gie 0.2, and the mass approximately traces light on large scales.
\end{abstract}

\section{Introduction} 
Theoretical arguments based on standard models of inflation, as well as 
on the demand of no
``fine tuning'' of cosmological parameters, predict a flat universe with the
critical density needed to just halt its expansion. 
The critical density, $\rm 1.9 \times 10^{-29} h^{2}g\ cm^{-3}$
(where h refers to Hubble's constant, see below), is equivalent to 
$\sim$10  protons
per cubic meter; this density provides the gravitational pull needed to slow down
the universal expansion and will eventually bring it to a halt.
So far, however, only a
small fraction of the critical density has been detected, even when all the
unseen dark matter in galaxy halos and clusters of galaxies is included.
There is no reliable indication so far that most of the matter needed to
close the universe does in fact exist. Here we show that several
independent observations of clusters of galaxies all indicate that the mass
density of the universe is sub-critical.  These observations include the mass
and mass-to-light ratio of clusters and superclusters of galaxies, the high
baryon fraction observed in clusters, and the evolution of the number density 
of massive clusters with time; the latter method provides a powerful measure not only 
of the mass-density of the universe but also the amplitude of the mass fluctuations.
The three independent methods-- all simple and robust-- yield consistent
results of a low-density universe with mass approximately tracing light on large
scales.

\section{Cluster Dynamics and the Mass-to-Light Ratio} 
Rich clusters of galaxies are the most massive virialized objects known.
Cluster masses can be directly and reliably determined using three
independent methods: (i) the motion (velocity
dispersion) of galaxies within clusters reflect the dynamical cluster mass,
within a given radius, assuming the clusters are in hydrostatic
equilibrium [1--3]; (ii) the temperature of the hot intracluster gas, like the
galaxy motion, traces the cluster mass [4--6]; and (iii) gravitational lensing
distortions of background galaxies can be used to directly measure the
intervening cluster mass that causes the distortions [7--10]. All three
independent methods yield consistent cluster masses (typically within radii
of $\sim$1 Mpc), indicating that we 
can reliably determine cluster masses within
the observed scatter ($\sim\pm$ 30\%).

The simplest argument for a low density universe is based on summing up
all the observed mass(associated with light to the largest possible scales) by 
utilizing the well-determined masses of clusters.  The masses of rich 
clusters of galaxies range from 
\gi10$^{14}$ to 10$^{15}$ h$^{-1}$M$_{\odot}$ within 1.5h$^{-1}$ Mpc radius of the cluster center (where $\rm h =  H_{0}/100\ km\ s^{-1}\ Mpc^{-1}$
denotes Hubble's constant).  When normalized by the cluster
luminosity, a median mass-to-light ratio of $\rm M/L_{B} \simeq 300 \pm 100 h$ in 
solar units ($\rm M_{\odot}/L_{\odot}$) is observed
for rich clusters, independent of the cluster luminosity, velocity
dispersion, or other parameters [3,11]. 
($\rm L_{B}$ is the total luminosity of the
cluster in the blue band, corrected for internal and Galactic absorption.) 
When integrated over the entire observed luminosity density of the universe,
this mass-to-light ratio
yields a mass density of $\rm \rho_{m} \simeq 0.4 \times 10^{-29} h^{2} g\ cm^{-3}$, or a mass density 
ratio of $\Omega_{m} = \rho_{m}/\rho_{crit} \simeq 0.2 \pm 0.1$
 (where $\rho_{crit}$ is the critical density needed to close the
universe). The inferred density assumes that all galaxies exhibit the same
high M/$\rm L_{B}$ ratio as clusters, and that mass follows light on large scales.
Thus, even if all galaxies have as much mass per unit luminosity as do massive
clusters, the total mass of the universe is only $\sim$20\% of the critical density. If
one insists on esthetic grounds that the universe has a critical density
($\Omega_{m}=1$), then most of the mass of the universe has to be unassociated with
galaxies (i.e., with light). On large scales (\ss 1.5\  h$^{-1}$ Mpc) the mass has to
reside in ``voids'' where there is no light. This would imply, for $\Omega_{m}=1$, a
large bias in the distribution of mass versus light, with mass distributed
considerably more diffusely than light.

Is there a strong bias in the universe, with most of the dark matter
residing on large scales, well beyond galaxies and clusters? An 
analysis of the mass-to-light ratio of galaxies, groups, and clusters
by Bahcall, Lubin and Dorman [11]
suggests that there is not a large bias. The study shows that the M/$\rm L_{B}$ ratio
of galaxies increases with scale up to radii of R \gi0.2  h$^{-1}$ Mpc, due to very
large dark halos around galaxies [see also 12,13]. The M/L ratio, however, appears to flatten
and remain approximately constant for groups and rich clusters from scales
of \gi0.2 to at least 1.5h$^{-1}$ Mpc and possibly even beyond (Figure 1). The 
flattening occurs at \mlb \gie\ 200--300h, corresponding to $\Omega_{m} \simeq 0.2$. 
(An \mlb\ \gie\ 1350 h is needed for a critical density universe, \gW=1.)
This observation contradicts the classical belief that the relative amount of dark matter
increases continuously with scale, possibly reaching \gW=1 on large scales. The
available data suggest that most of the dark matter may be associated with
very large dark halos of galaxies and that clusters do not contain a
substantial amount of additional dark matter, other than that associated
with (or torn-off from) the galaxy halos, plus the hot intracluster gas.
This flattening of M/L with scale, if confirmed by further larger-scale
observations, suggests that the relative amount of dark matter does not
increase significantly with scale above \gi0.2h$^{-1}$ Mpc. 
In that case, the
mass density of the universe is low, \gW\gi0.2--0.3, 
with no significant bias (i.e., mass approximately following light on large scales).

Recently the mass and mass-to-light ratio of a supercluster
of galaxies, on a scale of \gi6h$^{-1}$ Mpc, was directly measured using observations 
of weak gravitational lensing distortion of background
galaxies (Kaiser \gE\ [14]).  The results yield a supercluster mass-to-light ratio (on 6h$^{-1}$ Mpc
scale) of \mlb = 280 \gm 40h, comparable to the mean value observed for the three
individual clusters that are members of this supercluster.  These results provide
a powerful confirmation of the suggested flattening of \mlb\ (R) presented in Figure 1 (Bahcall \gE\
[11,15]).  The recent results confirm that no significant amount of 
additional dark matter exists on large scales.  The results also provide a clear
illustration that mass approximately traces light on large scales and that \gW\ is low,
as suggested by Figure 1.

\section{Baryons in Clusters} 
Clusters contain many baryons, observed  as gas and
stars. Within $\rm 1.5 h^{-1}$ Mpc of a rich cluster, the X-ray emitting gas
contributes \gi6h$^{-1.5}\%$ of the cluster virial mass [16--18].
Stars contribute another \gi2\%. 
The baryon fraction observed in clusters is thus:
\begin{equation}
\Omega_{b}/\Omega_{m} \gtrsim 0.06\rm h^{-1.5} + 0.02
\end{equation}
The observed value represents a lower-limit to the baryon fraction since we count only
the known baryons observed in gas and stars; additional baryonic matter may of course exist in the 
clusters.  Standard Big Bang nucleosynthesis limits the baryon density of the universe to [19,20]: 
\begin{equation}
\Omega_{b} \simeq 0.015 \rm h^{-2}
\end{equation}
These facts suggest that the baryon fraction observed in rich clusters (eq. 1) exceeds that of an \gW=1 universe ($\Omega_{b}/(\Omega_{m}$=1)\gie0.015h$^{-2}$; eq. 2)
by a factor of \ss3 (for h\ss0.5).  Since detailed hydrodynamic simulations [16,18] show that baryons
do not segregate into rich clusters, the above results imply that either the mean density
of the universe is lower than the critical density by a factor of \ss 3,
or that the baryon density is much larger than predicted by nucleosynthesis.
The observed high baryonic mass fraction in clusters (eq. 1), combined with the
nucleosynthesis limit (eq. 2), suggest (for h\gie0.65\gm0.1):
\begin{equation}
\Omega_{m} \lesssim 0.3.
\end{equation}
This upper limit on \gW\ is a simple, model--independent and thus powerful constraint: a critical
density universe is inconsistent with the high baryon fraction observed in clusters (assuming the
nucleosynthesis limit on $\Omega_b$, eq. 2).  [A universe dominated by hot dark matter that is
too hot to clump into clusters may survive this critical test; however, such a hot universe is 
unable to form the observed galaxies at high redshift, and is therefore unlikely.]

\section{Evolution of Cluster Abundance}
The observed present-day abundance of rich clusters of galaxies places a
strong constraint on cosmology: \gs\gW$^{0.5}$\gie 0.5, where \gs\ is the {\em rms} mass
fluctuations on 8h$^{-1}$ Mpc scale, and \gW\ is the present cosmological density 
parameter [21--26]. This constraint is degenerate in \gW\ and \gs; models 
with \gW =1, \gs\gi 0.5 are indistinguishable from models with \gW \gi  0.25, \gs \gi 1.
(A \gs \gie 1 universe is unbiased, with mass following light on large scales since galaxies 
(light) exhibits \gs(galaxies) \gie 1; \gs \gie 0.5 implies a biased universe with mass 
distributed more diffusely than light).

The {\em evolution} of cluster abundance with redshift, especially for massive
clusters, breaks the degeneracy between \gW\ and \gs\
[see, e.g. 23,24,27--34]. The evolution of high mass clusters is strong in 
\gW =1, low-\gs\ (biased) Gaussian models, where only a very low cluster 
abundance is expected at $z > 0.5$. Conversely, the evolution rate in low-\gW\, 
high-\gs\ models is mild and the cluster abundance at $z > 0.5$ is much higher than in 
\gW=1 models.

In low-density models, density fluctuations evolve and freeze out at early 
times, thus producing only relatively little evolution at recent times $(z\lesssim1)$.  
In an \gW=1 universe, the fluctuations start growing more recently thereby 
producing strong evolution in recent times; a large increase in the abundance of 
massive clusters is expected from $z\sim1$ to $z\sim0$.  In a recent study by 
Bahcall, Fan, and Cen [31] we show that the 
evolution is so strong in \gW=1 models that finding even a
few Coma-like clusters at $z > 0.5$ over \gi$10^{3}\ \rm deg^{2}$ 
of sky contradicts an \gW=1 model where only \gi$10^{-2}$ such clusters 
would be expected (when normalized
to the present-day cluster abundance). The evolution of the number
density of Coma-like clusters was determined from observations using the CNOC 
cluster sample to $z < 0.5$ and
compared with cosmological simulations [30--32]. The data show only a slow
evolution of the cluster abundance to $z \sim 0.5$, with \gi$10^{2}$ times more
clusters observed at these redshifts than expected for \gW=1. The results
yield \gW\gie 0.3\gm 0.1.

The evolutionary effects increase with cluster mass and with redshift. The
existence of the three most massive clusters observed so far at $z \sim 0.5-0.9$
places the strongest constraint yet on \gW\ and $\sigma_{8}$. These clusters
(MS0016+16 at $z=0.55$, MS0451$-$03 at $z=0.54$, and MS1054--03 at $z=0.83$, 
from the Extended Medium Sensitivity Survey, EMSS [35,36]), are nearly 
twice as massive as the Coma cluster, and have 
reliably measured masses (including gravitational lensing masses, 
temperatures, and velocity dispersions; [34,37--40].
These clusters posses the highest masses (\ss8 $\times 10^{14}$ h$^{-1}$ M$_{\odot}$ within 1.5 h$^{-1}$ comoving Mpc radius), 
the highest velocity dispersions (\ss1200 km s$^{-1}$), and
the highest temperatures (\ss8 keV) in the $z>0.5$ EMSS survey. 
The existence of these three massive distant clusters, even just the existence of the
single observed cluster at $z=0.83$, rules out Gaussian \gW=1 models for
which only \gi10$^{-5}\ z$ \gi\ 0.8 clusters are expected instead of the 1 cluster
observed (or \gi10$^{-3}\ z > 0.5$ clusters expected instead of the 3 observed). 
(See Bahcall \& Fan [34]).

In Figure 2 we compare the observed versus expected evolution of the number
density of such massive clusters. The expected evolution is based on the
Press-Schechter [41] formalism; it is presented for different
\gW\ values (each with the appropriate normalization \gs\ that satisfies the observed present-day cluster abundance, \gs\gie 0.5 \gW$^{-0.5}$; [23,26]).  The model curves 
range from \gW=0.1 (\gs\gie 1.7) at the top of the figure (flattest,
nearly no evolution) to \gW=1 (\gs\gie 0.5) at the bottom (steepest,
strongest evolution). The difference between high and low \gW\ models is
dramatic for these high mass clusters: \gW=1 models predict \gi10$^5$ times
less clusters at $z \sim 0.8$ than do \gW\gi 0.2 models.
The large magnitude of the effect is due to the fact that these are very massive
clusters, on the exponential tail of the cluster mass function;
they are rare events and the evolution of their number density depends exponentially on 
their ``rarity'', i.e., depends exponentially on $\sigma_{8}^{-2}\propto\Omega_m$ 
[32,34].  The number of clusters
observed  at $z \sim 0.8$ is consistent with \gW\gi 0.2, 
and is highly inconsistent with the \gi10$^{-5}$ clusters expected if 
\gW=1. The data exhibit only a slow, relatively
flat evolution; this is expected only in low-\gW\ models. 
\gW=1 models have a \gi10$^{-5}$ probability of producing the one 
observed cluster at $z \sim 0.8$, and, independently, a \gi10$^{-6}$ 
probability of producing the two observed clusters at $z \sim 0.55$. These 
results rule out \gW=1 Gaussian models at a
very high confidence level. The results are similar for models with or
without a cosmological constant. The data provide powerful constraints on \gW\ and $\sigma_{8}$: \gW=0.2$^{+0.15}_{-0.1}$ and $\sigma_{8} = 1.2 \pm 0.3$ (68\% 
confidence level) [34].  The high $\sigma_{8}$ value for the mean 
mass fluctuations indicates a nearly unbiased universe, with mass
approximately tracing light on large scales.  This conclusion is consistent with
the suggested flattening of the observed M/L ratio on large scales (Figure 1, section 2).

The rate of evolution of the cluster abundance depends strongly on \gs, as was shown 
by Fan \gE\ [32]: dlogn/dz $\propto$--$\sigma_8^{-2}$; low--\gs\ models evolve 
exponentially faster than high--\gs\ models, for a given mass cluster.  The strong 
exponential dependence arises because clusters of a given mass represent rarer 
density peaks in low--\gs\ models.  We show [32] that the {\it evolution rate} 
at z$\lesssim$1 is relatively insensitive to the density parameter \gW\ or to the 
exact shape of the power spectrum.  This is illustrated in Figure 3, where we present 
the cluster abundance ratio, n(z\gie 0.8)/n(z\gie 0), as a function of \gs\ for all \gW\ 
values.  The strong exponential evolution rate for low--\gs\ is clearly distinguished 
from the nearly no evolution expected for \gs\gie 1 (for any \gW).  The dependence 
on \gW\ is only secondary.  This method thus provides a unique tool for determining 
\gs --- by simply measuring the observed ratio of cluster abundance at different 
redshifts.  When combined with the observed normalization of the present-day cluster 
abundance, \gs\gW$^{0.5} $\gie 0.5, the density parameter \gW\ can be determined.  The 
observed evolution rate implied by the existence of the massive high redshift clusters 
discussed above is shown in Figure 3--- it indicates only minimal evolution thus a high 
\gs\ value of \gi 1.

In Figure 4 we summarize the above results for \gW\ and \gs\ as determined from 
cluster evolution and compare the results with the other two independent methods 
discussed in sections 2 and 3.  We show in Figure 4 the \gW$-$\gs\ band allowed 
by the present-day cluster abundance \gW$^{0.5}$\gie 0.5 \gs$^{-1}$, and the 
band allowed by the existence of the high-redshift massive clusters [34].  
The intersection of these two bands provides the cluster 
evolution constraints discussed above; the only allowed range for Gaussian 
models is low--\gW, \gs\gi 1.  We also present the \gW\ constraints determined 
from the two entirely independent methods discussed in \S\S 2--3: the high baryon fraction observed in 
clusters, which yields \gW\ls 0.3 (assuming the nucleosynthesis limit 
for the baryon density); and cluster dynamics which yields \gW\ \gie 0.2.  All these independent methods 
yield consistent results: \gW\gie 0.2 \gm 0.1 and \gs\gie 1.2\gm 0.2 (1 $\sigma$ level).  \gW=1 models are highly incompatible with these results.

\section{Summary} 
We have shown that several independent observations of clusters of galaxies all
indicate that the mass-density of the universe is sub-critical:
\gW\gie0.2\gm0.1.  The results also suggest that mass approximately traces light
on large scales.  A summary of the results, presented in Figure 4, is highlighted
below.

  1. The mass-to-light ratio of clusters (and superclusters) of galaxies and the suggested
     flattening of the mass-to-light ratio on large scales suggest \gW\gie0.2\gm 0.1.

  2. The high baryon fraction observed in clusters of galaxies suggests \gW\ls 0.3.

  3. The weak evolution of the observed cluster abundance to $z \sim 1$ provides a robust estimate of \gW\gie0.2$^{+0.15}_{-0.1}$, valid for any Gaussian models. An \gW=1 Gaussian universe is ruled
     out as a \ls10$^{-6}$ probability by the cluster evolution results (Figure 2--4).

  4. All the above-described independent measures are consistent with each
     other and indicate a low-density universe with \gW\gie0.2\gm0.1 (Figure 4). 
\gW=1 models are ruled out by the data. While non-Gaussian initial fluctuations, if they exist, will affect the cluster evolution results, they will not affect arguments (1) and (2) above. 
     Gaussian low-density models (with or without a cosmological constant)
     can consistently explain all the independent observations presented here.  These independent cluster observations indicate that we live in a lightweight 
universe with only \gi20\%--30\% of the critical density.   
Thus, the universe may expand forever.

\section{Acknowledgments}
I am grateful to X. Fan and R. Cen with whom much of the work reviewed here was done.  This work was supported in part by NSF grant AST93-15368.

\section{References}
\begin{enumerate}
\item  Zwicky, F., ``Morphological Astronomy" (Berlin: Springer-Verlag 1957).

\item  Bahcall, N.A., Ann. Rev. Astron. Astrophys. {\bf 15}, 505 (1977).

\item  Carlberg, R.G. \gE, Astrophys. J. {\bf 462}, 32 (1996).

\item  Jones, C. \& Forman, W., Astrophys. J. {\bf 276}, 38 (1984).

\item  Sarazin, C.L., Rev. Mod. Phys. {\bf 58}, 1 (1986).

\item  Evrards, A.E., Metzler, C.A., \& Navarro, J.F., Astrophys. J. {\bf 469}, 494 (1996).

\item  Tyson, J.A., Wenk, R.A., \& Valdes, F. Astrophys. J. {\bf 349}, L1 (1990).

\item  Kaiser, N. \& Squires, G. Astrophys. J. {\bf 404}, 441 (1993).

\item  Smail, I., Ellis, R.S., Fitchett, M.J., \& Edge, A.C., Mon. Not. Roy. Astron. Soc. {\bf 273}, 277 (1995).

\item  Colley, W.N., Tyson, J.A., \& Turner, E.L, Astrophys. J. {\bf 461}, L83 (1996).

\item  Bahcall, N.A., Lubin, L., \& Dorman, V., Astrophys. J. {\bf 447}, L81 (1995).

\item  Ostriker, J.P., Peebles, P.J.E., \& Yahil, A., Astrophys. J. {\bf 193}, L1 (1974).

\item  Rubin, V.C., Proc. Natl. Acad. Sci. USA {\bf 90}, 4814 (1993).

\item  Kaiser, N., \gE, astro-ph/9809268 (1998).

\item  Bahcall, N.A. \& Fan, X., Proc. Nat. Academy of Sciences, USA {\bf 95}, 5956 (1998).

\item  White, S.D.M., Navarro, J.F., Evrard, A., \& Frenk, C.S., Nature {\bf 366}, 429 (1993).

\item  White, D. \& Fabian, A., Mon. Not. Roy. Astron. Soc. {\bf 272}, 72 (1995).

\item  Lubin, L., Cen, R., Bahcall, N.A., \& Ostriker, J.P., Astrophys. J. {\bf 460}, 10 (1996).

\item  Walker, T.P., \gE, Astrophys. J. {\bf 376}, 51 (1991).

\item  Tytler, D., Fan, X.-M., Burles, S., Nature {\bf 381}, 207 (1996).

\item  Bahcall, N.A. \& Cen, R., Astrophys. J. {\bf 398}, L81 (1992).

\item  White, S.D.M., Efstathiou, G., \& Frenk, C.S., Mon. Not. Roy. Astron. Soc. {\bf 262}, 1023 (1993).

\item  Eke, V.R., Cole, S., \& Frenk, C.S., Mon. Not. Roy. Astron. Soc. {\bf 282}, 263 (1996).

\item  Viana, P.P. \& Liddle, A.R., Mon. Not. Roy. Astron. Soc. {\bf 281}, 323 (1996).

\item  Kitayama, T. \& Suto, Y., Astrophys. J. {\bf 469}, 480 (1996).

\item  Pen, U.-L., Astrophys. J. {\bf 498}, 60 (1998).

\item  Peebles, P.J.E., Daly, R.A., \& Juszkiewicz, R., Astrophys. J. {\bf 347}, 563 (1989).

\item  Oukbir, J. \& Blanchard, A., Astron. Astrophys. {\bf 262}, L21 (1992).

\item  Oukbir, J. \& Blanchard, A., Astron. Astrophys. {\bf 317}, 1 (1997).

\item  Carlberg, R.G., Morris, S.M., Yee, H.K.C., \& Ellingson, E., Astrophys. J. {\bf 479}, L19 (1997).

\item  Bahcall, N.A., Fan, X., \& Cen, R., Astrophys. J. {\bf 485}, L53 (1997).

\item  Fan, X., Bahcall, N.A., \& Cen, R., Astrophys. J. {\bf 490}, L123 (1997).

\item  Henry, J.P., Astrophys. J. {\bf 489}, L1 (1997).

\item  Bahcall, N.A. \& Fan, X., Astrophys. J. {\bf 504}, 1 (1998).

\item  Henry, J.P, \gE, Astrophys. J. {\bf 386}, 408 (1992).

\item  Luppino, G.A. \& Gioia, I.M., Astrophys. J. {\bf 445}, L77 (1995).

\item  Smail, I., Ellis, R.S., Fitchett, M.J., \& Edge, A.C., Mon. Not. Roy. Astron. Soc. {\bf 273}, 277 (1995).

\item  Luppino, G.A. \& Kaiser, N., Astrophys. J. {\bf 475}, 20 (1997).

\item  Mushotsky, R. \& Scharf, C.A., Astrophys. J. {\bf 482}, L13 (1997).

\item  Donahue, M., \gE, Astrophys. J. {\bf 502}, 550 (1998).

\item  Press, W.H. \& Schechter, P., Astrophys. J. {\bf 187}, 425 (1974).
\end{enumerate}
\newpage

\begin{figure}
\centerline{\vbox{
\psfig{figure=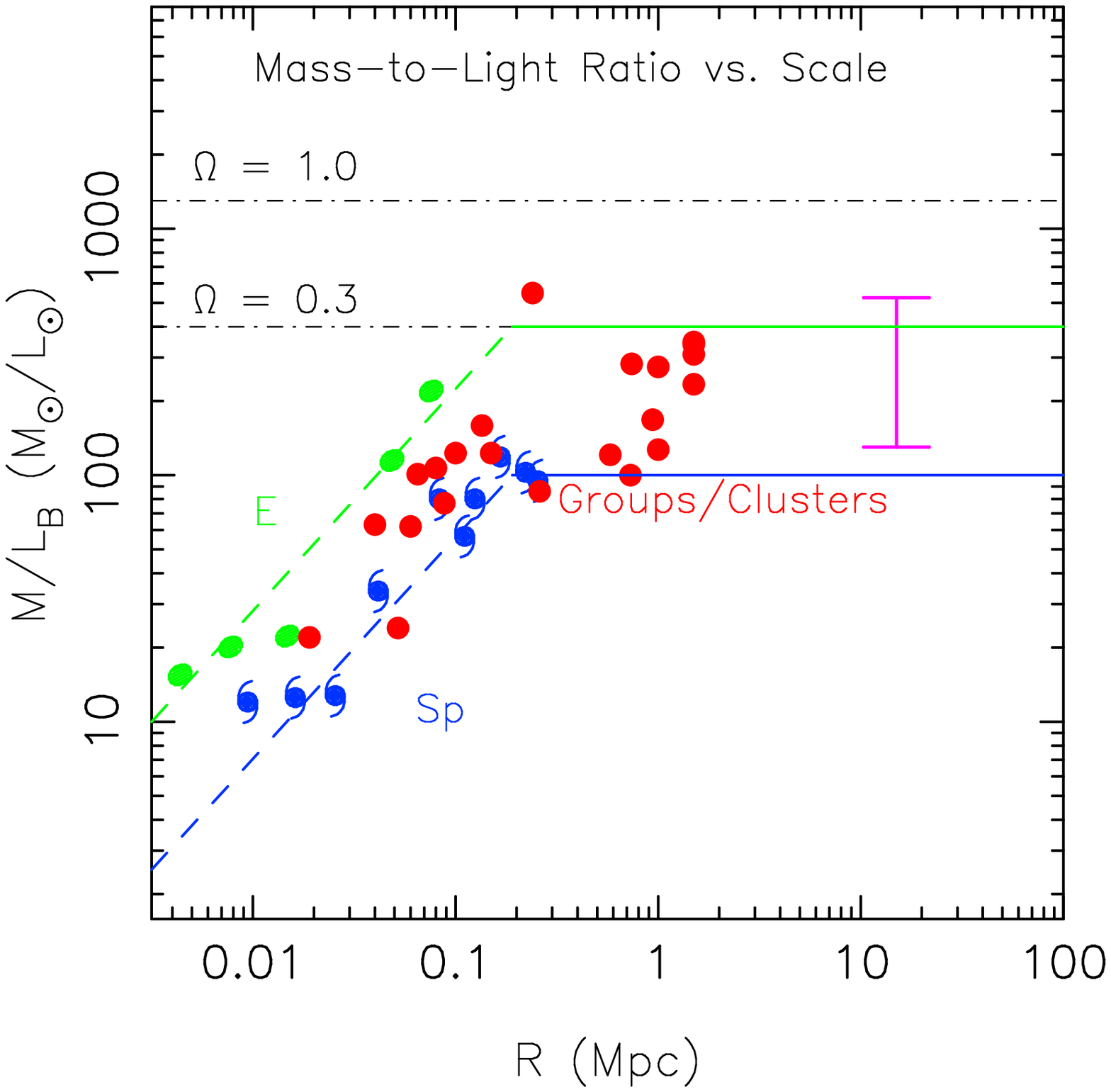,height=5in}}}
\caption[{\footnotesize Figure 1.}]{\footnotesize The dependence of mass-to-light ratio,
\mlb, on scale, R, for average spiral galaxies (spiral symbols), elliptical galaxies (elliptical symbols), and groups and clusters (filled circles).  Adapted from Bahcall, Lubin and Dorman [11]; [15].  The large scale point at \gi15 h$^{-1}$ Mpc represents Virgo cluster infall motion results [11].  The location of $\Omega_{m}=1$ and $\Omega_{m}=0.3$ are indicated by the horizontal lines. A flattening of \mlb\ is suggested at $\Omega_{m} \simeq 0.2 \pm 0.1$.  A recent 
result for a supercluster using weak gravitational lensing (R\gie6h$^{-1}$Mpc; [14]), \mlb=280\gm40h, is consistent with the suggested flattening of M/L(R).}  
\end{figure}
\newpage

\begin{figure}
\centerline{\vbox{
\psfig{figure=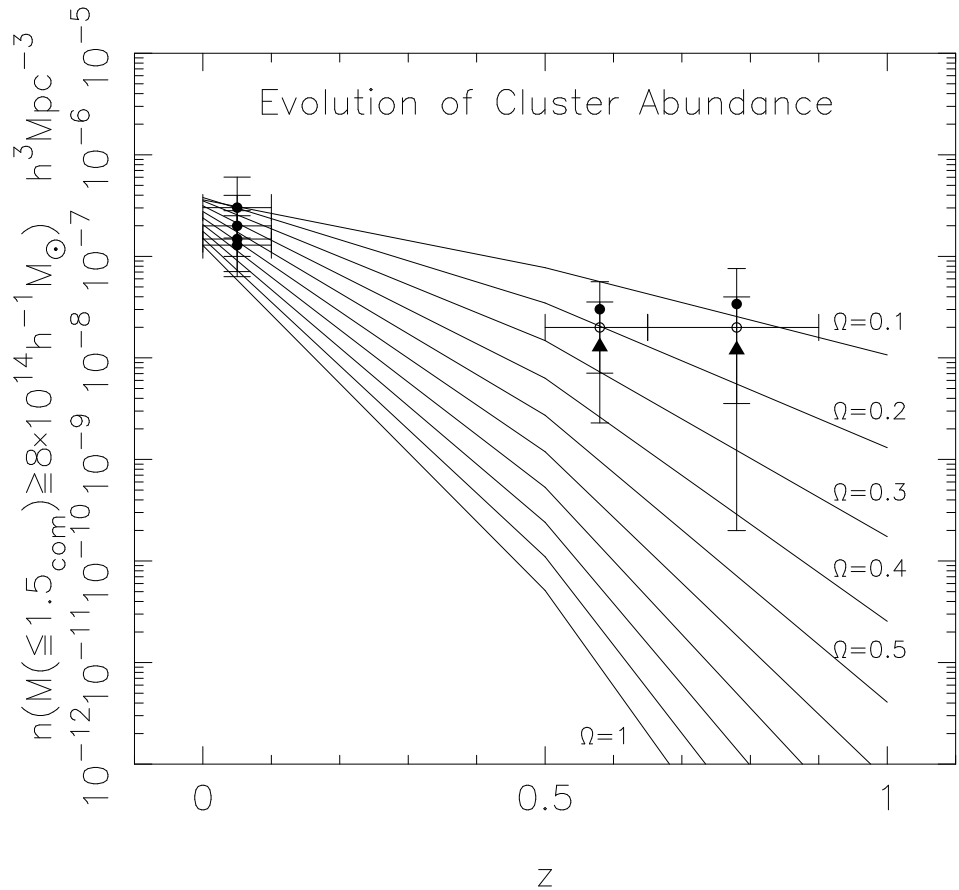,height=5in}}}
\caption[{\footnotesize Figure 2.}]{\footnotesize Evolution of the number density of massive
clusters as a function of redshift: observed versus expected (for clusters with mass \ss 8 $\times$ 10$^{14}$ h$^{-1}$ M$_{\odot}$ within a comoving radius of $\rm 1.5 h^{-1}$ Mpc).  (Adapted from Bahcall and Fan 1998 [34].  The expected evolution is presented for different
\gW\ values by the different curves.  The observational data points (see text) show only a slow evolution in the cluster abundance, consistent with \gW\ \gie 0.2$^{+0.15}_{-0.1}$.  Models with \gW=1 predict \gi 10$^{5}$ fewer clusters than observed at $z$ \gi 0.8, and \gi 10$^{3}$ fewer clusters than observed at $z$ \gi 0.6.}
\end{figure}
\newpage

\begin{figure}
\centerline{\vbox{
\psfig{figure=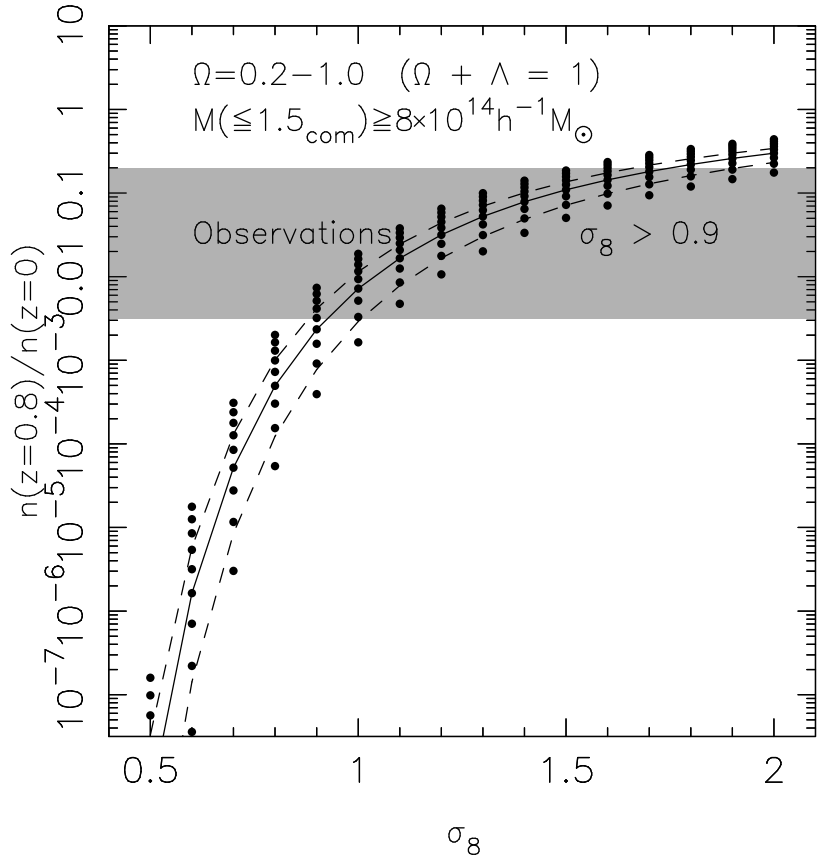,height=5in}}}
\caption[{\footnotesize Figure 3.}]{\footnotesize Cluster~abundance ratio, $n(z=0.8)/n(z=0)$,
vs. \gs\ from Press-Schechter (solid curve, for mean of all \gW's) for clusters with mass \ss8 $\times$ 10$^{14}$h$^{-1}$ M$_{\odot}$.  Filled circles represent \gW's from 0.2 to 1 (bottom to top).  (Dashed curves represent the mass threshold range of 7 to 10 $\times$ 10$^{14}$ M$_{\odot}$, top and bottom, respectively).  The data (Figure 2, \S 4) are shown by the shaded region (68\% level).  Similar results are obtained for $\Lambda=0$.}
\end{figure}
\newpage

\begin{figure}
\centerline{\vbox{
\psfig{figure=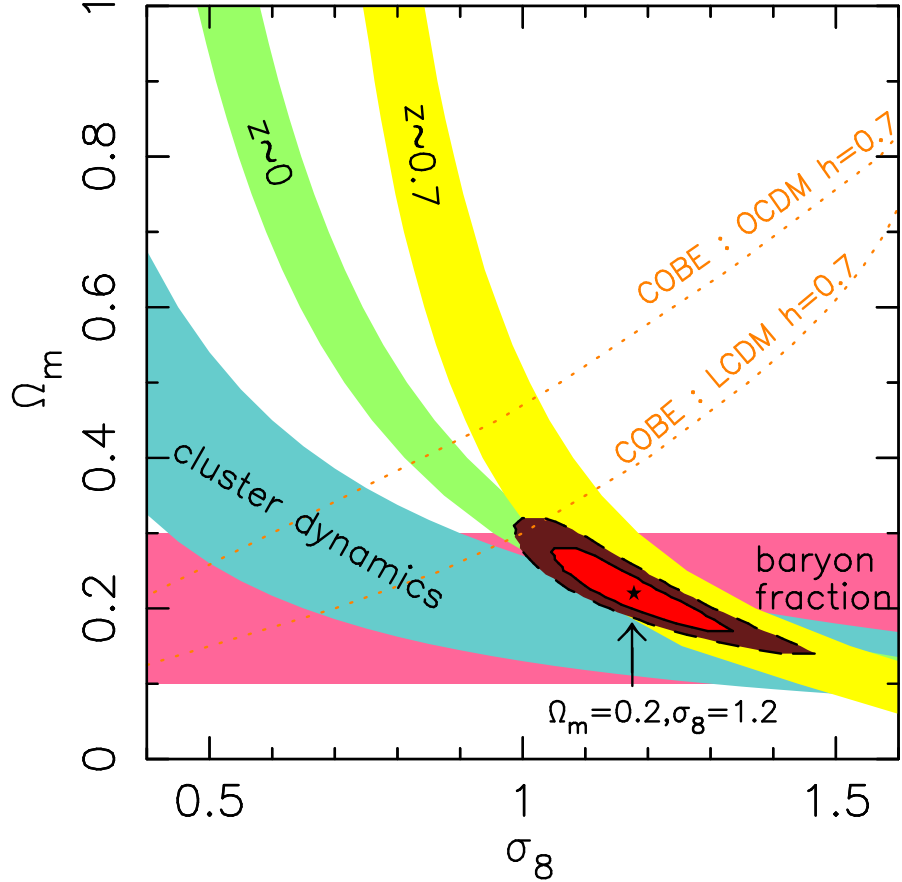,height=5in}}}
\caption[{\footnotesize Figure 4.}]{\footnotesize Constraining the mass-density parameter,
$\Omega_{m}$, and the mass fluctuations on $\rm 8 h^{-1}$ Mpc scale, $\sigma_{8}$,
from several independent observations of clusters: cluster dynamics; baryon fraction in clusters; present-day cluster abundance ($z \sim 0 $); and cluster abundance at redshift $z \sim 0.7$.  (The latter two abundances yield the cluster evolution constraints shown in Figure 2; see text).  All these model-independent observations converge at the allowed range of $\Omega_{m} = 0.2 \pm 0.1$ and $\sigma_{8} = 1.2 \pm 0.2$ ( 68\% confidence level).  The dotted lines illustrate the mean microwave fluctuations constraints, based on the COBE satellite results, for a Cold-Dark-Matter model with h = 0.7 (with and without a cosmological constant, denoted as LCDM and OCDM respectively. Both models are consistent, within their uncertainties, with the best-fit $\Omega_{m} - \sigma_{8}$ regime of the cluster observations).}
\end{figure}

\end{document}